# Quantifying the Deign Quality of Object Oriented System
# The metric based rules and heuristic

R.Selvarani
Associate Professor,
Dayananda Sagar College of Engg.
Bangalore. Karnataka

Dr. Wahida Banu
Prof. & Head ECE Dept
Govt. College of Engg.
Salem- Tamil Nadu.

Dr. Kamakshi Prasad
Professor- CSE Department
JNT University
Hyderabad- Andhra Pradesh

**Abstract**

The design structure of OO software has decisive impact on its quality. The design must be strongly correlated with quality characteristics like analyzability, changeability, stability and testability, which are important for maintaining the system. But due to the diversity and complexity of the design properties of OO system e.g. Polymorphism, encapsulation, coupling it becomes cumbersome.

Maintenance is recognized as the most difficult and expensive activity of software development process. (It takes 70% of production cost.) In order to support maintenance of an OO software system, the quality of its design must be evaluated using adequate quantification means.

To help developers and maintainers to detect and localize design flaws in a system, in addition to the metrics, we propose a novel mechanism called Detection strategy for formulating metric based rules that capture deviations from good design principles and heuristics.

We have defined detection strategies for capturing possible design flaws of OO design and validated the approach experimentally on multiple large scale case studies.

**Key criteria of good Object Oriented Design:**

1. Low coupling
2. High cohesion
3. manageable complexity
4. Proper data abstraction

**OO design metrics:**

In 1994 Chidember and kemerer proposed a now widely accepted suite of metrics for an oo system. Basili validated the metric suite in 1996 and Tang in 1999. The six metrics are,

1. Weighted methods per class (WMC): It measures the complexity of the individual class.
2. Depth of inheritance trees (DIT): It is defined as the length of the longest path of inheritance ending at the current module.
3. Number of Children (NOC): It represents the number of immediate subclasses subordinated to a class in the class hierarchy.
4. Coupling between objects (CBO): It is defined as the count of the number of other classes to which it is coupled.
5. Response for a class (RFC): It gives the number of methods that can potentially be executed in response to a message received by an object of that class.
6. Lack of cohesion in methods (LCOM): It counts for number of method pairs whose similarity is zero minus the count of method pairs whose sililarity is not zero.

**Possible design flaws in OO design:**

**Improper coupling:** Class level-------------Shotgun Surgery.
Subsystem level----- Wide subsystem interface.

**Low cohesion:** Method level--------- Feature Envy
Subsystem level-------Misplaced class

**Improper distribution of complexity:**
Class level--------------God class.
Method level------------God method.
Subsystem level----------God package.

**Flaws related to data abstraction:**
Class level-----------------Data class, Refused Bequest.

**Micro Design level flaws:** Lack of bridge, Lack of Strategy, Lack of state, Lack of singleton, Lack of Facade.

**Detection Strategy:**
A detection strategy is the ***quantifiable expression of a rule*** *by which design fragments that are conforming to that rule can be detected in the source code*.

It is therefore a generic mechanism for analyzing a source code model using metrics. "*Quantifiable expression of a rule*" means that the rule must be properly expressible using object- oriented design metrics.

The use of metrics in the detection strategies is based on the mechanisms of **filtering** and **composition.**

**The Filtering Mechanism:**
Data filtering is to reduce the initial data set to detect those design fragments that have special properties captured by the metric. The limits (margins) of the subset are defined based on the type of data filter. Extreme (abnormal) values or values that are in a particular range will be considered normally for data filtering. Therefore we identify two types of filters.

- *Marginal Filter*- a data filter in which one extremity (margin) of the result set is implicitly identified with the corresponding limit of the initial data set.
- *Interval Filter*- a data filter in which both the lower and upper limit of the resulting subset are explicitly specified in the definition of the data set.

**Marginal Filters:**
It depends on the limit(s) specified. Marginal filters are of two types.

1. Semantically: It requires two parameters, They are
    a. Threshold value- Indicates the marginal values.
    b. Direction- Indicates the marginal values.

2. Statistical: It does not need any explicit specification for threshold because it is determined directly from the initial data set by using statistical methods.
    E.g. Box plot, Standard deviation.

But the direction must be specified.

Types of Semantical Filters:

1. Absolute Semantical Filters: HigherThan and LowerThan are used to express the sharp design rules or heuristics. – e.g. a class should not be coupled with more than 6 other classes.
2. Relative Semantical Filter: Top Values and Bottom Values can be specified are relative to the initial data set.

Interval Filters: Here we need to specify two threshold values. Example
Between (2 0, 3 0) can be composed out of two (semantical) marginal filters i.e. HigherThan (200 and LowerThan (30).
In addition to the filtering mechanism we also have a **composition mechanism,** interval filters in all cases reducible to a composition of two marginal filters of opposite directions.

**Composition Mechanisms:**
It supports correlated interpretation of multiple result sets. There are three operators **and, or, butnot**. It can be analyzed in two view points.

1. Logical view point: The **‘and’** operator suggests the coexistence of both the symptom described on the left side of it as well as the existence of the symptom presented on the right side.

2. Set view point: the **’and’** operator will mapped to intersection, ‘**or**’ operator to reunion and the **‘but not’** operator to set minus.

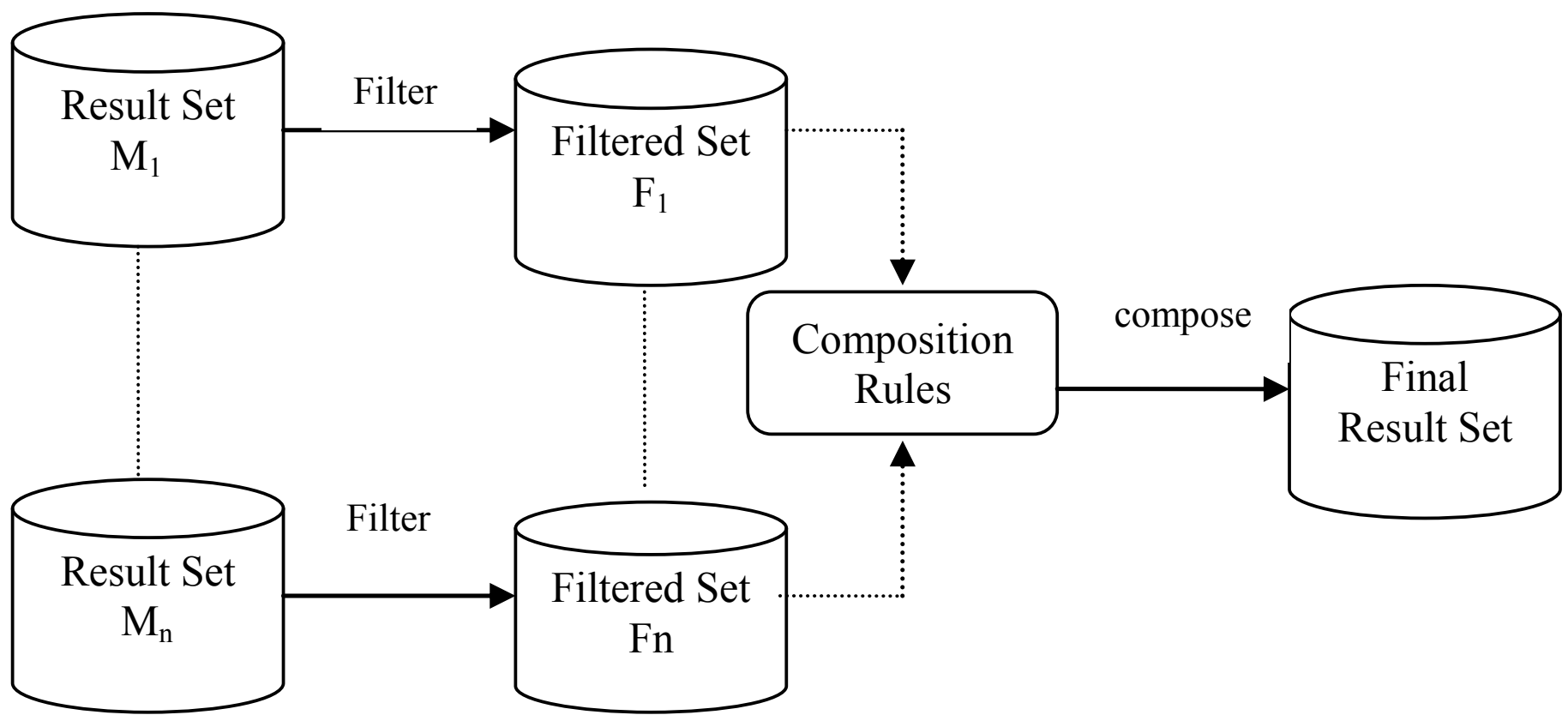

**The filtering and composition from a set view point**

**Selecting the Data Filters:**

The following are the rules to choose a data filter for a particular metric, while quantifying the design rules or heuristics.

***Rule 1:*** Choose an absolute semantical filter when quantifying design rules that specify explicitly concrete threshold values.

***Rule 2:*** Choose a relative semantical filter when the design rule is defined in terms of fuzzy marginal values, like “the high/low values” or “the highest/lowest values”.

***Rule 3:*** For large systems, parameterize relative semantical using percentile values. On the other hand, use absolute parameters when applying a relative semantical filter to small-scale systems.

***Rule 4:*** Choose a statistical filter for those cases where the design rules make reference to extremely high/low values, without specifying any precise threshold.

**Accuracy of detection strategy:**
It depends upon the threshold values used in parameterizing any detection strategy. This can be improved by using any of the following mechanism.

1. **Experience and hints from literature:** In most of the cases setting the threshold values is highly empirical process and it is guided by similar past experiences and by hints from metrics' authors.
2. **Tuning Machine:** A promising approach is found in [25] where the author defines a "tuning machine**"** which tries to find automatically the proper threshold values, and thus tune the detection strategies. This approach is based on building a repository of flaw samples. Based on this reference samples, the threshold values are fixed which maximize the number of correctly detected samples. The drawback of the approach is that the examples repository must be large enough, which is bit complicated.
3. **Analysis of Multiple Versions:** History of the system is considered for fixing the threshold. Detection process can be enhanced by combining detection strategies applied on a single version with additional information about the history of the system (i.e. Version analysis).
4. **Defining a Detection Strategy:** Consider a design flaw **"behavioral God Class."**
   Consider the three heuristics found in Riel's book.

   a. Top level classes in a design should work uniformly.
   b. Beware of classes with much non communicative behavior.
   c. Beware of classes that access directly data from other classes.

**Step wise methodology:**

1. Implement above heuristics suggested to our example. (**"behavioral God Class.")**

   a. Uniform distribution of intelligence among classes which refers to 'high class complexity'
   b. Intra class communication refers to 'low cohesion of classes
   c. Special type of coupling i.e. direct access to instance variable refers to 'access to foreign data'.

2. Selection of proper metrics that quantify the identified properties: the selected example, God class is related to class complexity, cohesion of classes and access to foreign data, which are best properties to be quantified. Metrics can be selected based on the above.

   Weighted Method Count (WMC): It is sum of statistical complexity of all methods.
   Tight Class Cohesion (TCC): It is the relative number of directly connected methods.

Access to foreign data (AFTD): It represents the number of external classes from which a given access the attributes directly or via accessor methods.

3. To find suitable filtering mechanism:
   For the first symptom "high class complexity" 'the top values' relative semantical filter is chosen for WMC metric.
   For the second 'low cohesion' symptom the 'bottom value' relative semantic filter for TCC metric.
   For the third 'to find any access to foreign data' 'higher than' absolute filter is used.

4. Threshold Setting:
   Consider a 50% value for both the 'top values' filter attached to the metric WMC and 'Bottom values' filter attached to the metric TCC. For AFTD, no direct access to the data of other classes is permitted, so the threshold value is 1.

5. Correlate the se symptoms: These three symptoms are correlated using composition operator 'and'. This connects all symptoms and the detection strategy for God classes is formally described in the following equation.

   God class(s) = (WMC(C), TopValues (50%)) ~ (AFTD(C),HigherThan(1)) ~ (TCC,BottomValues(50%))

**Implementation of Detection Strategies:**

1. Express the detection strategies in a computer understandable format. A language deals only with the filtering and composition of metric results and relying on SQL queries for the computation of individual metrics can be used. Ex: SOD (System object design suggested by C.Chiril, timisora univ.)
2. The SOD file will be having the detection strategies with distinct name for easy computation to find critical design fragments.

**Inspection Process:**

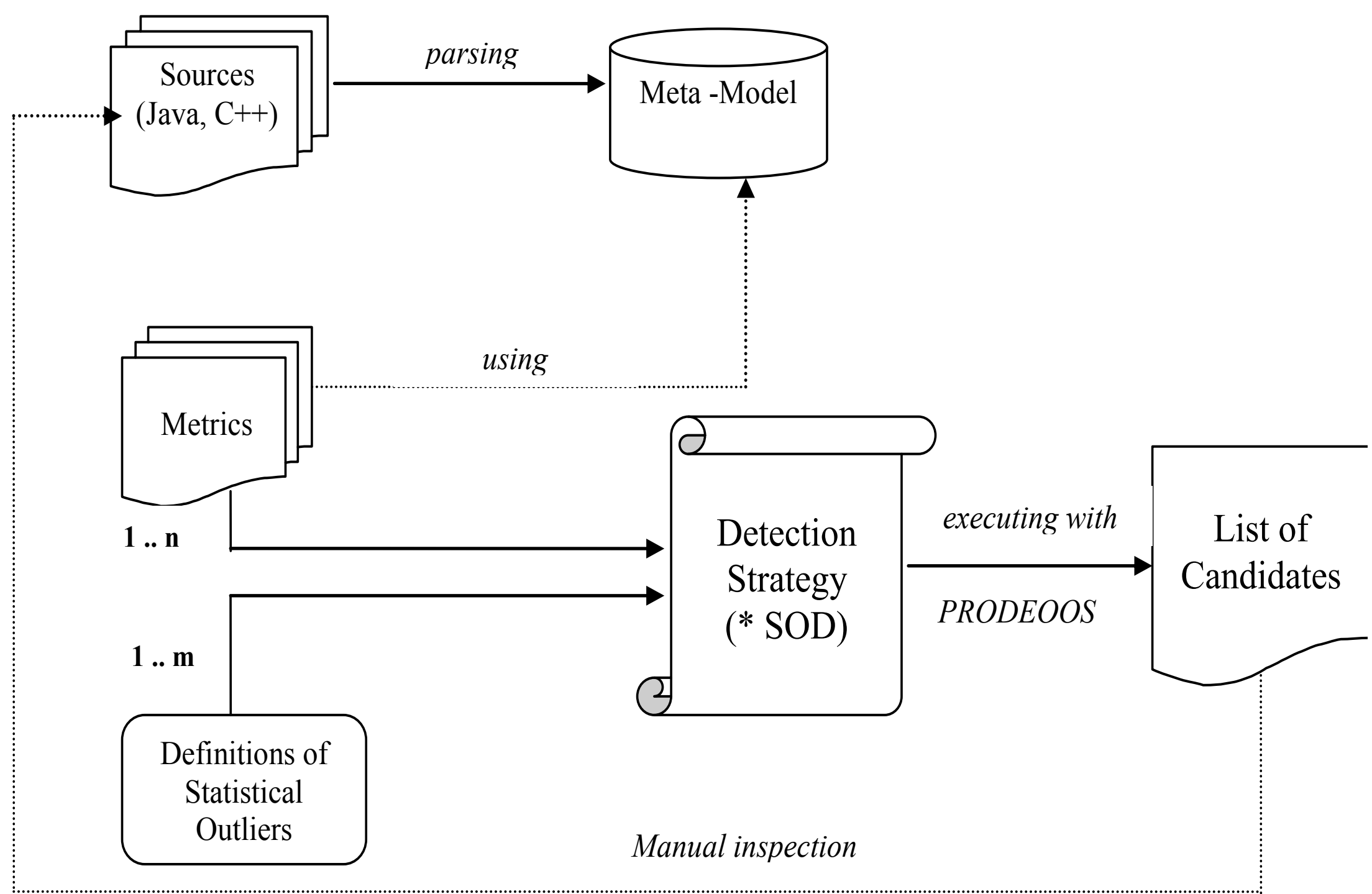

In order to apply metric system on a given software system, we need to have all necessary design information which is stored in meta model consists of information about the design entities(e.g. classes, methods, variables) of the system and about the exiting relations(e.g. inheritance, call of methods) among these entities. This model is used as an abstraction layer for various object oriented programming languages.

The detection strategy implemented as a SOD script , can be automatically run using the PRODEOOS tool on the design model of the system to be analyzed. The result is a set of design entities together (e.g. classes, methods) that are reported at suspect for that particular detection strategies. PRODEOOS returns the values for the different metrics that were involved in the detection strategy. The results obtained in the previous step must be manually inspected in order to decide if the suspects conform to the rule quantified in the detection strategy.